\documentclass[12pt]{emulateapj}
\usepackage{hyperref}
\usepackage{graphicx}
\usepackage{amssymb}
\usepackage{epstopdf}
\usepackage{times}
\usepackage{natbib}
\def\deg{\circ}
\def\fermi{Fermi {}}

\def\be{\begin{equation}}
\def\ee{\end{equation}}
\shorttitle{Galactic center at very high-energies}
\shortauthors{Chernyakova et al.}

\begin{document}
\title{Galactic center at very high-energies.}
\author{Chernyakova, M.$^1$, Malyshev, D.$^1$, Aharonian, F.~A.$^{1,2}$, Crocker, R.~M.$^2$, Jones, D~I.$^2$}
\email{masha@cp.dias.ie}

\affil{$^1$ Dublin Institute for Advanced Study, Astronomy \& Astrophysics Section, 31 Fitzwilliam Place, Dublin, 2 Ireland}
\affil{$^2$ Max Planck Institut f\"{u}r Kernphysik, Postfach 103980, 69029 Heidelberg, Germany}

\begin{abstract}
Employing data collected during the first 25 months' observations by the \fermi-LAT,
we describe and subsequently seek to model the very high energy ($>300$~MeV) emission 
from the central few parsecs of our Galaxy.  
We analyse, in particular, the morphological, spectral and temporal characteristics of the central source, 1FGL~J1745.6-2900. 
 Remarkably,  the data show a clear, statistically significant signal at energies above 
10 GeV, where the  \fermi-LAT  has an excellent angular resolution comparable to the angular resolution of HESS
at TeV energies. This  not only reduces dramatically the contamination both from the  diffuse background  and 
the nearby gamma-ray sources,  but also makes meaningful the joint analysis of the Fermi and HESS data. 
Our analysis  does not show statistically significant   variability of 1FGL~J1745.6-2900.   
Using the combination of \fermi data on 1FGL~J1745.6-2900 and HESS data on the coincident, TeV source 
HESS~J1745-290, we show that the spectrum of the central $\gamma$-ray source is inflected with a relatively steep spectral region matching between the  flatter spectrum found at both low and high energies.
We seek to model the  gamma-ray production in the inner 10~pc of the Galaxy 
and
examine, in particular,  cosmic ray (CR) proton propagation scenarios that reproduce 
the observed spectrum of the central source.
We show that a model that instantiates a transition from diffusive propagation of the  CR protons at low energy to almost rectilinear propagation at high energies (given a reasonable energy-dependence of the assumed diffusion coefficient) can well explain the spectral phenomenology.
In general, however, we find considerable degeneracy between different parameter choices which will only be broken with the addition of morphological information that $\gamma$-ray telescopes cannot deliver given current angular resolution limits.
We  argue  that a future analysis done in combination with higher-resolution radio continuum data holds out the promise of 
breaking this degeneracy.

\end{abstract}


\section{Introduction}
Over the past decade-and-a-half since the discovery by EGRET of a  very high energy (VHE) gamma-ray source near the Galactic center (GC), there has been intense speculation as to what mechanism(s) are producing the observed emission.
The subsequent  discovery of TeV gamma-ray emission from the Sgr A* region by the ground-based
gamma-ray instruments, in  particular by the HESS array of atmospheric Cherenkov telescopes
\citep{Aharonian04},  has generated further theoretical activity.
Of general interest  and import -- given the GC constitutes  the nearest example of a galactic nucleus --
is the question  concerning the sites and mechanism(s) by which particles are accelerated to
TeV energies and beyond in the dynamical center of our Galaxy.

Despite the fact that the GC TeV gamma-ray   source  is a  point-like  object  for HESS,
the $0.07^\circ$ PSF of the instrument and the extremely crowded and complex nature of the
region (as evidenced by the complex radio morphology \citep{Law2008})
do not allow the unambiguous identification of the source(s) of  gamma-ray emission.
With the latest data, however, it is possible to place the center-of-gravity of the TeV point source within the central $\sim6''$ of the Galaxy \citep{hess_gc10}, leaving only a handful of possible sources.
These include the central black hole itself, Sgr~A* \citep{AN05_apj,Liu2006}; a plerion discovered within the central few arcseconds of the Galaxy \citep{wang06,Hinton07},  a putative  "black hole plerion"   produced by the wind
from Sgr~A*  \citep{atoyan_dermer04},  and  the diffuse   $\leq 10$~pc  region
surrounding  Sgr~A* \citep{AN05_apss,Ballantyne07,Ballantyne10}.

Given the above background,
we consider here the further insights now possible
in light of the  \fermi-LAT observations of the GC region.
In particular, since the PSF of Fermi above $\sim10$~GeV is similar  to that of HESS,
it is possible to explore   a quite broad  energy interval   of relativistic  particles localized
in this region.
%

First 11 months of \fermi observations of the GC were presented by J.Cohen-Tanugi on behalf of \fermi LAT collaboration during the 2009 Fermi Symposium.  In this work, the  authors argued  that the \fermi source 1FGL~J1745.6-2900 and the HESS source  J1745-290 are spatially coincident.  Also, they derived  an energy  spectrum of the \fermi source
till 100 GeV and  concluded  that to match the HESS spectrum either a high-energy break or a cut-off is required.

In this work we analyze 25 months of \fermi LAT data.
In addition to the central GeV source and other  reported sources,
our analysis  reveals  four new sources of GeV gamma-rays  located  in this region.
With the spectral information from both Fermi and HESS in hand, we model below the production of
gamma-rays from the inner GC due to hadronic interactions of protons accelerated within the central
black hole  and diffusing into the surrounding interstellar medium.

In section \ref{sec:analysis}, we describe the reduction and analysis of the \fermi data. We present the details of our model in section \ref{sec:modeling}. In section \ref{sec:discussion} we discuss the implications
 of the obtained results   and  summarize the main conclusions  in section \ref{sec:summary}.

\section{Data Analysis and Results.}\label{sec:analysis}
The large area telescope (LAT) on board the \fermi satellite is a pair-conversion gamma-ray detector operating between 20~MeV and 300~GeV. The LAT has a wide field of view of $\sim$2.4 sr at 1~GeV, and observes the entire sky every two orbits ($\sim$3 hr for a \fermi orbit at an altitude of $\sim$565 km; full details of the instrumentation are given in \cite{Atwood09}). The data used for our analysis  are based on the
first 25 months of observations (August 4, 2008 -- August 18, 2010).

The data analysis was performed using the LAT Science Tools package with the \texttt{P6\_V3} post-launch instrument response function \citep{Rando09}. The standard event selection for source analysis, resulting in the strongest background-rejection power (diffuse event class) was applied. In addition, photons coming from zenith angles larger than $105^\deg$ were rejected to reduce the background from gamma rays produced in the atmosphere of the Earth. The analysis was further restricted to the energy range above 100 MeV, because below this energy effective area becomes very small and the residual uncertainty in the instrumental response is significant.

\begin{figure*}
\includegraphics[width=0.33\textwidth]{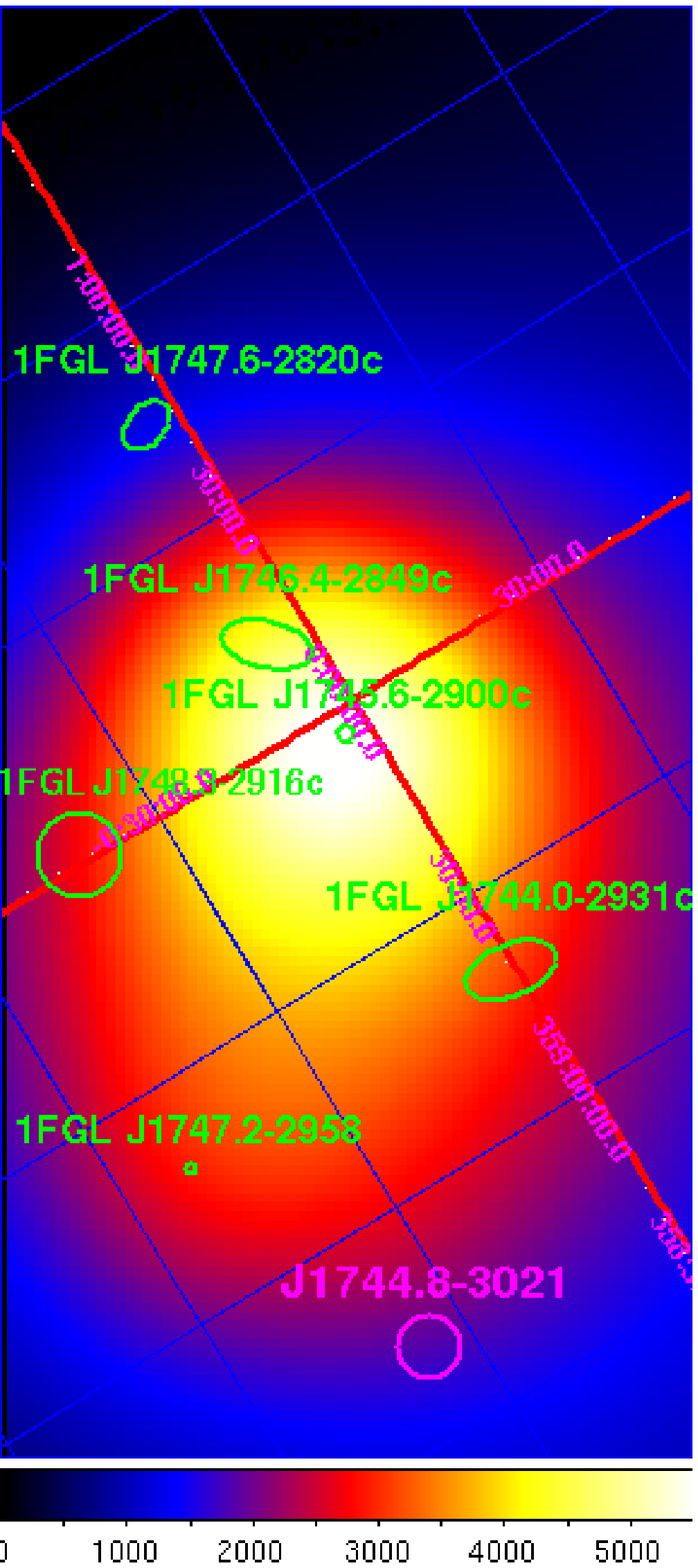}
\includegraphics[width=0.33\textwidth]{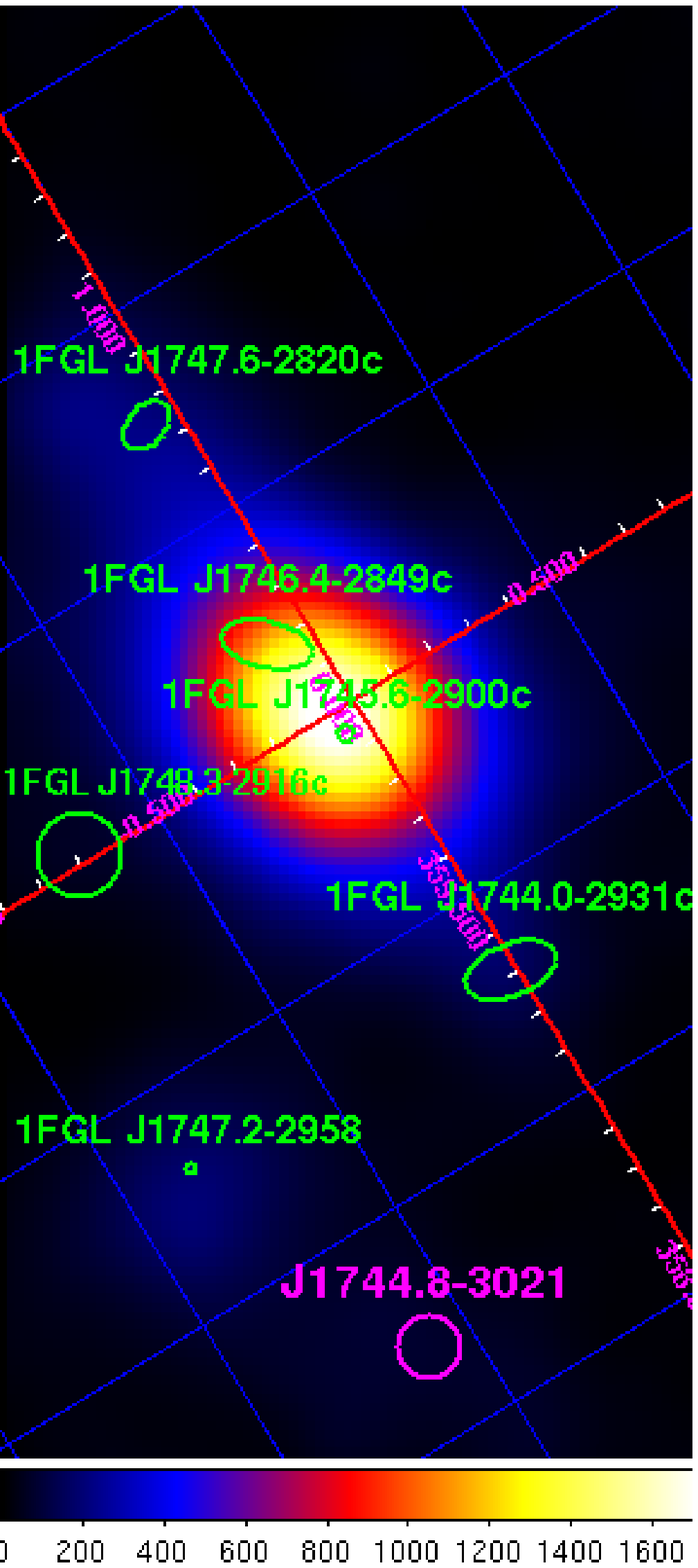}
\includegraphics[width=0.33\textwidth]{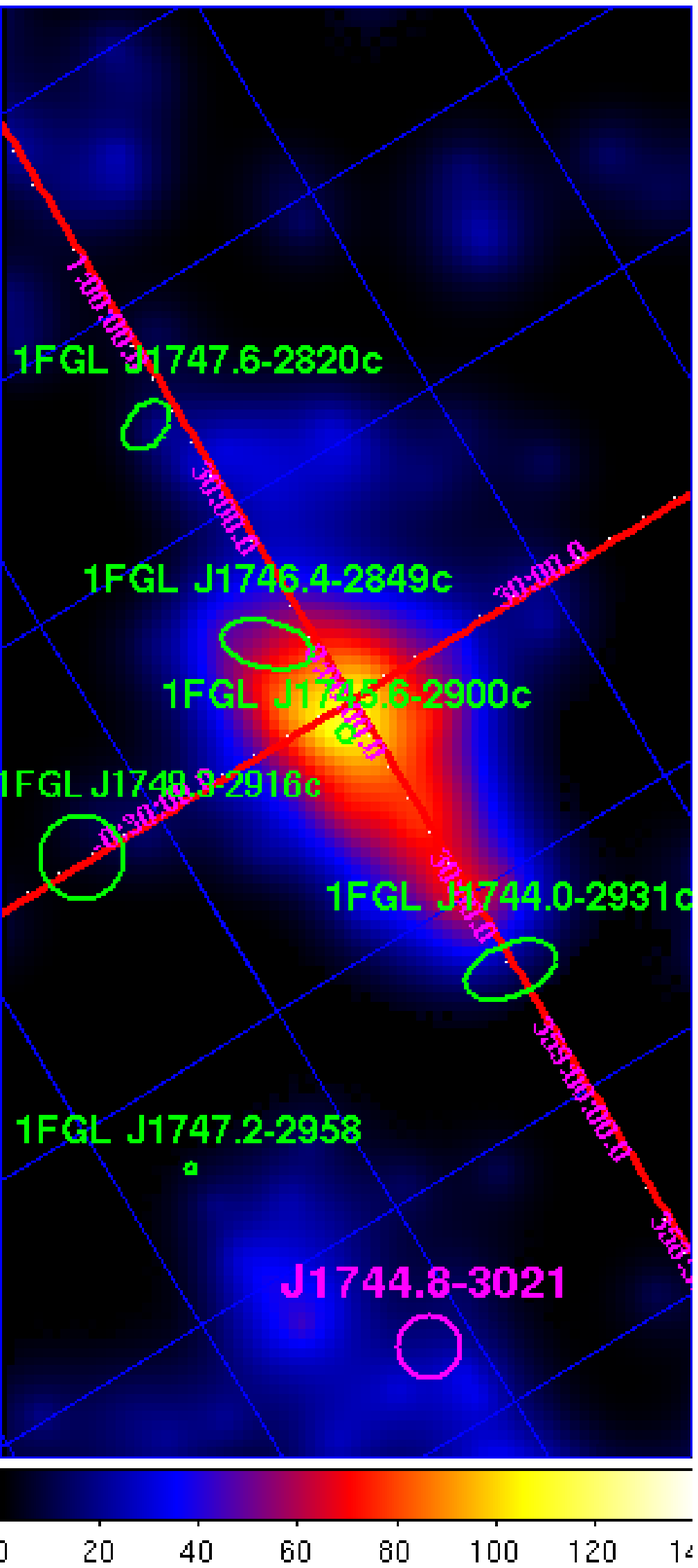}
\caption{TS maps of the central part ($1.5^\deg \times 3.5^\deg$) of the Galaxy center
as seen by \fermi in 300 MeV -- 3GeV, 3GeV--30 GeV and  30GeV--300 GeV energy
ranges (left to right).
Positions of new sources are marked with magenta circles.
Green ellipses correspond to the positions of the sources from
the 1 year \fermi catalogue. Note that linear colour scheme has different maximum value in all cases varying from 5500 in the less energetic  left picture to 140 in the most energetic right one. Source significance can be approximately estimated as a square root of TS.}
\label{gcmap}
\end{figure*}

In order to take into account the broad point spread function (PSF) at low ($\sim 100$~MeV) energies, we constructed a sequence of test statistic (TS) images of  the  $10^{\deg}\times10^{\deg}$ region around the Sgr~A*. In producing TS images, we used the \texttt{gttsmap} tool with a tolerance parameter of $ftol=10^{-5}$ and a bin size in each map of $0.1^{\deg}$. Finally, after subtracting the 19 known sources from the one year \fermi catalogue (1FGL) which happen to be within the selected region, we  found four  new sources, which are listed in the Table~\ref{newsrc}, in the residual images. One of these sources (indicated as J1744.8-3021) -- shown in magenta in Figure \ref{gcmap} -- lie within the $1.5^\deg \times 3^\deg$ area around the GC.
This source coincides spatially with known HESS source HESS J1745-303 and EGRET source 3EGJ1744-3011.

\begin{table}
\centering
\caption{Coordinates  and TS of the new sources discovered during the analysis.}
\label{newsrc}
\begin{tabular}{@{}ccc}
\hline
RA&Dec&TS\\
			(J2000.0)	&	(J2000.0)& \\	
\hline
 264.906&-28.555&331\\
 266.210&-30.360&424\\
 270.060&-30.091&189\\
 270.697&-30.626&192\\
\hline
\end{tabular}
\end{table}

\begin{figure}
\includegraphics[width=\columnwidth]{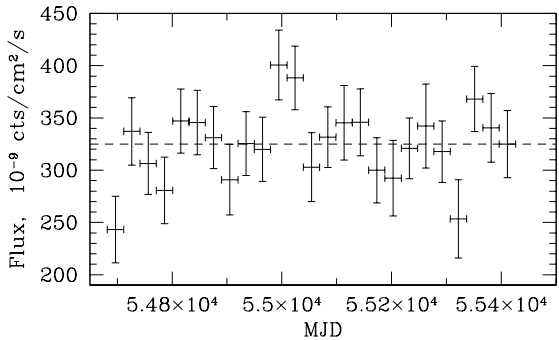}
\caption{Lightcurve of the 1FGL~J1745.6-2900 in 300 MeV -- 100 GeV energy range. The average flux is shown with a dashed line.}
\label{lc}
\end{figure}

In order to construct a light-curve for 1FGL~J1745.6-2900, we used a spectral method by selecting data in 300~MeV--100~GeV energy range and fitting all known sources, selected as above, with a single power law model. Afterwards, we split the whole time interval into 25 equal time bins and fit source spectra by fixing
their slopes to the best-fit value obtained over the entire time period, leaving the source normalization as a free parameter.
The normalization of the Galactic and extra-galactic background was also left as a free parameter. The resulting light-curve is shown in Figure \ref{lc} and is relatively stable and  does not show any statistically significant variation.
The averaged flux is equal to $(324.9 \pm 7.05)\times 10^{-9}$ counts~cm$^{-2}$~s$^{-1}$, with a reduced $\chi^2=1.1$ for 24 degrees of freedom.

Spectral fitting was performed within 100~MeV--300~GeV energy range with the \texttt{gtlike} tool. The spectrum in 100~MeV--300~GeV energy range can be fitted by a power law with a slope of $\Gamma=2.212 \pm 0.005$ and a flux normalization of $F=(1.39\pm0.02) \times 10^{-8}$~cm$^{-2}$~s$^{-1}$~MeV$^{-1}$ at 100~MeV. We also attempted to split the spectrum into two different energy bands, and found that the fitted slope is equal to $\Gamma=2.196 \pm 0.001$ in 300~MeV--5~GeV energy range, and $\Gamma=2.681 \pm 0.003$ in 5 -- 100 GeV energy range. The errors given above are statistical errors  and represent the 1$\sigma$ deviation.  Thus the slope of the \fermi spectrum above several GeV is  significantly steeper than the spectrum  reported by the HESS collaboration at TeV  energies ($F_{\mbox{HESS}}\sim E^{-2.1}$, \citep{GC09_spectra}. Note that at low energies, \fermi has a very broad PSF, rapidly moving from 4$^\deg$ at 100~MeV to 2$^\deg$ at 300~MeV. Thus, taking into account the possible source confusion in the region, one should treat the first point in the spectrum (100--300~MeV) with caution.

\section{Modeling}\label{sec:modeling}

As  proposed  in \cite{AN05_apss}, a significant fraction of the protons accelerated near the black hole may enter the surrounding gaseous environment and initiate VHE gamma-ray emission  through neutral pion production and subsequent  decay.
The efficiency of  the process, and the energy spectrum  of resulting gamma-rays
depends not only  on the protons' injection rate and  the ambient gas density,
but  also on  the speed of proton transport into the surrounding medium
\citep{AN05_apss,Ballantyne07,Ballantyne10}.
 To explain the  gamma-ray spectrum reported by the HESS collaboration, \cite{AN05_apss} assumed that relativistic protons with a power-law spectrum possessing a spectral index of $\Gamma \sim2$ are injected into the dense gaseous environment surrounding the central black hole. The diffusion coefficient, $D$, was assumed to have a power-law dependence on energy of the form $D(E)=10^{28}(E/1$GeV$)^\beta\kappa$~cm$^2$~s$^{-1}$.  For the cosmic ray
diffusion in the Galactic disk $\kappa \sim 1$ and $\beta \sim 1$, but of course
the diffusion coefficient in the GC could be quite different. In \cite{AN05_apss}  the
parameter  $\beta$  was assumed to be in the range of 0.5--1.
In the model of \cite{Ballantyne07,Ballantyne10}  the propagation is treated using the  ray-tracing technique.
They found that in order to reproduce
in their model the reported energy distribution of TeV  gamma-rays,
the  spectrum of protons should be hard with a spectral index   $\sim 0.75$.
Such an exceptionally  hard injection  spectrum of protons implies a very strong energy dependence
of the character of propagation of protons which, within the formalism of diffusion,
would require a diffusion  coefficient  with  $\beta \sim 1.5$.

Given that the VHE emission detected by HESS and \fermi can be localized to within the central several arcminutes then, for a GC distance of $d\sim$ 8~kpc, the linear size of the production region of VHE gamma-rays can be as large as 10~pc. Continuum X-ray and radio observations of the central 10~pc region of the Galaxy  show  a complex environment with many unique structures \citep{Law2008, Crocker2010}. For simplicity in this paper  we  assume that the supermassive black hole in the center of our Galaxy is surrounded by a shell of a dense matter with a density, normalized to $n_H=1000$ cm$^{-3}$ at 1 pc radius, with either constant or $1/r^2$ radial dependence.
The inner and outer radii of this shell are  parameters in our model.
Another relevant parameter is the time evolution of the  proton  injection.  Although one can treat it as
a quasi-stationary process, in fact the proton injection  can be dominated by one or several flares that occurred
in the past in Sgr~A*. In this context, one should mention  the morphological interpretation of the
diffuse gamma-ray emission observed by HESS from the
central 200~pc region of GC, which relates the  positive detections  of gamma-rays from
giant molecular clouds in GC
to a putative "proton" flare that occurred  in Sgr~A*  in the past,  10,000 years ago or so  \citep{Aharonian06_GC}.
The detection of  reflected X-ray emission from the Sgr B2 cloud  is another, more direct  piece of evidence about the short flaring activity of  Sgr~A* a  few hundred years ago \citep{sunyaev93,koyama96, koyama08, Revnivtsev04, Terrier10}.

In the standard diffusion approximation the propagation of particles is described by the diffusion equation \citep{ginzburg64} which, in the spherically symmetric case, reduces to the form:
\begin{equation}
 {\partial n \over \partial t} =  {D\over r^2} {\partial  \over \partial r} r^2{\partial n \over \partial r} + {\partial  \over \partial E} (bn) + Q,
\label{diffeq}
\end{equation}
where $n(r,t,E)$ is the space density of relativistic particles with energy $E$, at instant $t$ being a distance $r$ from the source; $b(e)=-dE/dt$ is the continuous energy loss rate;  $Q(E,t)$ is the injection rate; and $D(E)$ is the energy-dependent diffusion coefficient. We have assumed here, for simplicity, that $D$ is independent of $r$ and has a power-law dependence on energy as stated above. The solution of  equation (\ref{diffeq}) can be written as \citep{Syrovatskii59}:
\begin{equation}
n(E,r,t)= \int^t_0 P(E,r,t-x)Q(E,x)dx,
\end{equation}
where the propagator, $P(E,r,t)$, is defined as:
\begin{equation}\label{propclass}
P(E,r,t)=\frac{1}{[4\pi \lambda(E,t)]^{3/2}} \exp\left[-\frac{r^2}{4\lambda(E,t)} \right],
\end{equation}
and
\begin{equation}\label{lambda}
\lambda(E,t)=-\int_{E_g(t)}^E dx \frac{D(x)}{b(x)}.
\end{equation}
In equation (\ref{lambda}) $E_g$ is the energy that a cooled particle has at time $t$, if its initial energy was $E$.

 Formally,  the diffusion equation does not contain information on how fast a particle may propagate.
Since Eq.(\ref{diffeq})  does not prevent an artificial "superluminal motion" ($v_0=2D(E)/r \gtrsim c$),  we follow
the phenomenological approach proposed by \cite{berezinsky09} who
introduced  a propagator, $P(E,r,t)$ in the form:
\begin{equation}
P(E,r,t)=\frac{\theta(1-\xi)}{4\pi (ct)^3} \frac{1}{(1-\xi^2)^2}\frac{\alpha(E,\xi)}{K_1[\alpha(E,\xi)]} \exp \left[ -\frac{\alpha(E,\xi)}{\sqrt{1-\xi^2}} \right],
\label{propber}
\end{equation}
where $\theta(x)$ is the Heaviside step function, $\xi(t)=r/ct$, $K_1(x)$ is modified Bessel function of the second kind, and $\alpha(E,t)$  is defined as:
\begin{equation}
\alpha(E,t)=\frac{c^2 t^2}{2 \lambda(E,t)}.
\end{equation}
 In the low-energy regime (i.e., $E\ll E_c$), the propagator  given by Eq.(\ref{propber}) reproduces the standard treatment  of diffusion, whilst in the high-energy regime ($E\gtrsim E_c$) it describes particles that move in a rectilinear fashion. Here $E_c$ is the energy at which  Eq. (\ref{diffeq}) allows  diffusion with the speed of light
\begin{equation}
E_c=\left(\frac{cR}{2D_0}\right)^{1/\beta} \mbox{GeV.}
\label{Ec}
\end{equation}

\begin{figure}
\begin{center}
\includegraphics[width=\columnwidth]{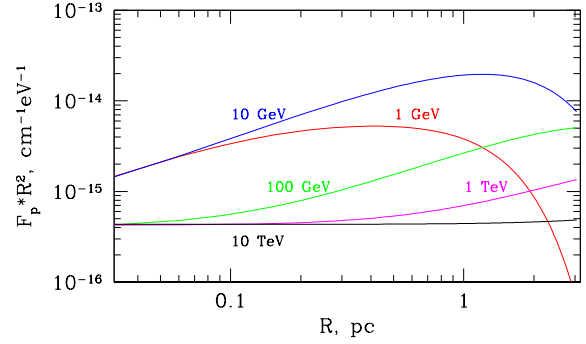}
\caption{ The fluxes  of protons as a function of radius at different energies shown as marked. For ease of comparison each energy was multiplied  by a factor of 900 (for  10 GeV), 8000 (100 GeV), $5\times 10^6$ (1TeV) and $7\times 10^8$ (10~TeV). The flux at each energy have been multiplied by $R^2$, so that rectilinear propagation corresponds to a horizontal line.\label{protprof}}
\end{center}
\end{figure}

Due to the energy dependence of the diffusion coefficient,
proton propagation will be quite different at low as compared to high energies.
We have explored how this plays out in the GC environment. The result  presented in Figure \ref{protprof}
shows the change of the radial distribution of protons as a function of energy.
We have determined the proton distribution after 300 years of continuous injection into the interstellar medium of density $n_H=10^3$cm$^{-3}$ within a region of radius $R=3$pc.  The initial spectrum of protons was assumed to have  a power-law distribution with an exponential cutoff, $Q(E)\propto E^{-2}\exp(-E/100$~ TeV$)$.

It can be clearly seen from Figure~\ref{protprof} that, whilst at 10 TeV the particles pass through the region in an almost rectilinear fashion, at lower energies, protons propagate diffusively. An explanation for the lack of low-energy protons at high radii can be found in the fact that these particles have too low an escape velocity  to travel that far in a given time.

\begin{figure}
\begin{center}
\includegraphics[width=\columnwidth]{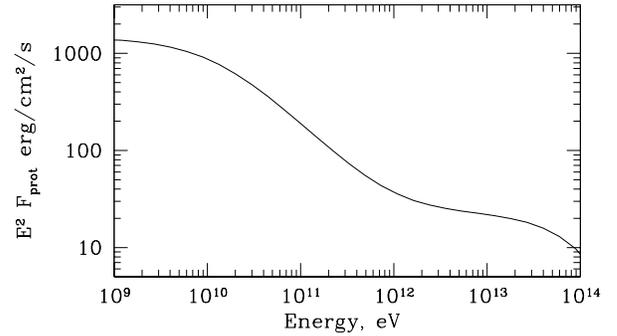}
\caption{ The  energy distribution of  protons averaged over 3pc of gamma-ray production region,
as reconstructed from \fermi and HESS data.\label{protspec}
}
\end{center}
\end{figure}

The spectrum of the protons integrated over the  gamma-ray production region (see  Figure \ref{protprof})
is shown in Figure \ref{protspec}. Photons produced by the interaction of relativistic protons with such an energy distribution fit both \fermi and HESS data well. At low (GeV) energies, the diffusion radius is smaller than the region so that protons are accumulated within the region
and, given the almost energy-independent $pp$ cross-section, mirror the spectrum of  the injected protons.
On the other hand, at TeV energies protons begin to propagate in a rectilinear mode and will have again the form of the injected spectrum, albeit at a lower   flux  level.
Protons with an intermediate energy have a much steeper, diffusion-processed spectrum representing the transition between the two extremes. The spectral shape of the highest energy gamma-rays is not affected by the propagation effects. Therefore
in order to match the spectrum at highest energies  reported by HESS \citep{GC09_spectra}, we assumed
an exponential cut-off in the proton spectrum and fix its position  at 100 TeV.


\begin{figure}
\begin{center}
\includegraphics[width=\columnwidth]{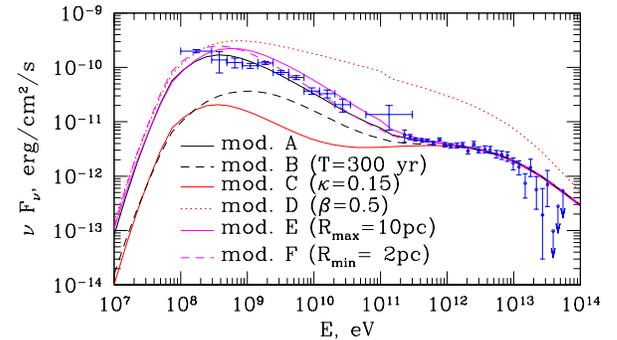}
\caption{Spectral energy distribution of gamma-rays expected from a region filled with relativistic and non-relativistic protons within different assumptions concerning the injection, diffusion and the region geometry (see text for a discussion of  parameters for each specific model). The  data points have been derived from the \fermi and HESS data\label{diffmod}}
\end{center}
\end{figure}

 Below, we fit parameters which represent the particle injection spectrum, the propagation of the injected protons, and the geometry of the interstellar medium. It is instructive to systematically examine the influence of these model parameters on the resulting spectrum. To do this we begin with $10^4$~years of injection of relativistic protons with a spectrum of the form $Q(E)\propto E^{-1.9}\exp(-E/100$~TeV$)$ into a 3~pc radius region filled by an interstellar  gas  of constant density, $n_H=10^3$~cm$^{-3}$. The injection rate was taken equal to $Q_0=3.9\times 10^{39}$~erg/sec. The diffusion parameters were chosen as $\beta=0.9$, $\kappa=0.015$ in order to reproduce the combined \fermi and HESS spectrum.   The photon spectrum resulting from this parameter set (model A)  is shown  in Figure \ref{diffmod}. The other curves in Figure \ref{diffmod} illustrate the effect of change of a single parameter, while all others are fixed to the values used in model A.

Figure \ref{diffmod}  corresponds to the case of a source active for only 300 years (model B).
This change does not  affect high energy particles, traveling rectilinearly, because their escape time is $t_{esc}=R/c \sim 10$~years, much shorter then the injection time.
These particles fully fill the region and their density  is   the same as in the case of model A.
The diffusion time at low energies ($E<E_c\sim$1 TeV in this case), however, is much longer (viz. $t_{diff}=R^2/2D\sim 9000$ years at 1GeV). Thus, 300 years will be not enough for low-energy particles to travel to the outer regions of the shell,  and the total spectrum  is expected to  be  harder with respect to model A. Thus, in this parameter set, whilst the radiation does not differ from our heuristic case at high energies, at lower energies there are necessarily fewer gamma-rays.

Models C and D in Figure \ref{diffmod} show what occurs when the diffusion parameters $\kappa$ and  $\beta$  are changed. If one increases the diffusion coefficient by a factor of 10, then by the same factor, the diffusion time of the low-energy particles is decreased, leading to a corresponding reduction in the intensity of the gamma-ray emission. If one changes the energy dependence of the diffusion by decreasing $\beta$ by a factor of two, then the transition of the particles propagation from diffusion to rectilinear propagation occurs at much higher energies ($E_c$ has $1/\beta$ dependance, see equation (\ref{Ec}), and reaches  PeV energy in this case). Thus the emission  increases at all energies and the spectral form changes due to a larger influence of high energy particles.

Models E and F in Figure \ref{diffmod} show the change of the photon spectrum due to the change of the geometry of the region.  The corresponding  curve in figure  corresponds to a shell geometry with an outer radius of $R=10$~pc (as compared to the 3~pc radius considered in the other models). Using this particular geometry causes the overall normalization of the resulting spectrum to increase (a factor of 0.25 was applied to this spectrum for easier comparison in the production of Figure \ref{diffmod}) due to the increase in size of the gamma-ray production region. Additionally, the `bump' at lower energies becomes wider, since the energy at which particles transition from the diffuse to the rectilinear propagation regimes increases as $R^{1/\beta}$, and thus the larger the radius, the larger the number of 'intermediate', GeV, energy particles that can be accumulated.

The effect of the shell volume is clearly seen in the case when the value of the inner radius of the disk is changed. Model F in Figure \ref{diffmod}  is for a shell with inner radius $R_{min}=2$pc and outer radius $R_{max}=3$pc. In this case the overall normalization of the resulting spectrum decreases (in Figure \ref{diffmod} a factor of 3.3 has been applied to this spectrum in order to more easily aid the comparison to other models), but the low-energy bump is still wider than in the case of a shell with no hole (i.e. an oblate spheroid). This is due to the fact that removing the inner part of the shell mostly diminishes the soft photon emission, effectively increasing the relative number of more energetic ones.

 Finally note that the sharp drop of gamma-ray spectra below 1 GeV is the result of the kinematics of pion production at p-p interactions, and therefore  does not depend  on the model parameters.

\begin{figure}
\begin{center}
\includegraphics[width=\columnwidth]{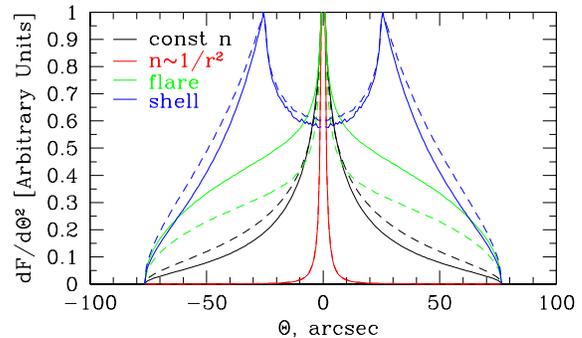}
\caption{Brightness profile of the GC in 100MeV - 1GeV (solid lines) and 1GeV -- 10GeV (dashed lines) energy ranges.
 See Table \ref{param} for  the description of the models.  25$''$ corresponds to 1pc.
\label{brightprof}}
\end{center}
\end{figure}

\begin{table}
\caption{Parameters of different models  used for Figure \ref{brightprof}.  The
medium density at R=1pc was assumed to be equal to $n_H=10^3$cm$^{-3}$.
$Q_0$ and $Q_f$ correspond to the injection rates of the constant source and a flare correspondingly. }
\begin{center}
\begin{tabular}{|c|c|c|c|c|c|c|}
\hline
model&n$_H$&$\kappa$&$\beta$&Q$_0$&R$_{min}$&Q$_f$\\
&&&&10$^{39}$erg/s&pc&10$^{39}$erg/s\\
\hline
const&const  &0.02 &0.95&6  &0&0\\
$1/r^2$&$1/r^2$&0.002&0.75&0.1&0&0\\
flare&const  &0.7  &0.5 &8  &0&8000\\
shell&const  &0.02 &0.95&8  &1&0\\
\hline
\end{tabular}
\end{center}
\label{param}
\end{table}

The radial distribution of photons is also highly dependent on the model parameters. Figure~\ref{brightprof} shows the brightness profile of the inner 3~pc after $10^4$~years of the constant injection at energies between 100~MeV--1~GeV (solid lines) and 1~GeV--10~GeV (dashed lines). For all models the constant source was active for 10$^4$ years. The initial spectrum of protons is assumed to be a power-law with an
exponential cutoff, $Q(E)\propto E^{-2}\exp(-E/100$~TeV$)$.   The
medium density at R=1pc was assumed to be equal to $n_H=10^3$cm$^{-3}$.At higher energies particles pass through the region almost rectilinearly and the region will appear as a point-like source of gamma-rays to \fermi.  As labeled in Figure \ref{brightprof}, different curves correspond to different model parameter sets in terms of the ambient matter distribution and injection. For all models in Figure \ref{brightprof}, parameters, given in Table \ref{param}, were chosen so that the resulting integrated emission accurately reproduces the \fermi and HESS data, and the resulted profiles were normalized to the maximum value to aid comparison.

Geometrically, the solid line represents the case of constant density and exhibits a broader profile at higher energies (line). If instead, one models the region with a density falling off proportional to a $n_H\propto 1/r^2$ profile, the resulting profiles are thinner and are represented by the red solid and dashed lines, which almost coincide on the figure. The profile is more centrally peaked in the latter case, since the matter is more concentrated in the center and so the photon flux will originate mostly from this region.

The green solid and dashed lines show the profile created if, in addition to a constant source, there was also a flare which occurred 300 years ago (the injection rate in the Table \ref{param} corresponds to a flare of length 10 years). In order to match observational data in this case, a larger diffusion coefficient had to be assumed, which inevitability leads to a larger diffusion radius and -- correspondingly -- to a wider profile. Finally in Figure \ref{brightprof} we show the brightness profiles corresponding to a shell geometry. In this case the profile has a maximum at a radius of the inner shell.


\section{Discussion}\label{sec:discussion}
The spectral properties of the very high energy emission from the GC differs considerably from that at lower energies: whilst the GC is known to be variable at X-rays and near-IR wavelengths, no variability has been detected either by HESS, or by Fermi. This seeming duality has a natural explanation if the low energy emission is generated very close to the central black hole, while  the gamma-ray emission originates from a much larger region and is emitted during the diffusion of the relativistic protons through the interstellar medium surrounding the central black hole.  In such a case, the very high energy emission would only reflect (with a delay) major flares originating from the central source.
As remarked above, one interpretation of the distribution of the diffuse, TeV $\gamma$-ray emission relative to the molecular clouds in the central $\sim200$~pc of our Galaxy, is that a central CR proton source flared about $10^4$~years ago \citep{Aharonian06_GC}.
In the previous section we showed that our model is able to reconstruct data if we assume a constant injection of relativistic protons for $1.0\times10^4$~years (see Figure \ref{diffmod}). This time is higher than the diffusion time for the used set of parameters and thus the obtained photon  energy spectrum  is effectively steady state.

 The observations of  reflected X-radiation from the cloud Sgr B2 located at a distance of 100~pc from Sgr~A*
suggests that a few hundred years ago there was  an increase of X-ray luminosity of Sgr~A*  \citep{sunyaev93,koyama96, koyama08, Revnivtsev04, Terrier10}. In our modeling, we checked that we are able to explain
the data as a result of a constant injection  of protons over three hundreds years.
We found that if one assumes an injection rate of protons of $6\times 10^{39}$ erg/s, and takes $\beta=0.95$, $\kappa=0.01$ (the resulting radial distribution of protons for this model is shown in Figure \ref{protprof}), then the resulting emission is in good agreement with the observations.
 Our model is able to self-consistently explain different spectral  indices at GeV and TeV energies by the different effective escape velocities of the protons.  While high energy protons, producing TeV photons, escape quasi-rectilinearly without spectral deformation, as, indeed, do the particles fully trapped at the lowest energies, the particles with intermediate energies are affected by diffusion, but not fully trapped and their spectrum becomes much steeper providing the transition between two extreme cases.

Recent monitoring of the Sgr~B2 cloud with X-ray instruments shows flux variability on time scales of 10~years \citep{Terrier10}.
This variability can be naturally interpreted as a measure of the flare duration.   Although the X-ray luminosity and proton acceleration in Sgr~A*  should not necessarily  correlate, it is interesting to explore also the scenario
when we deal with a flare of proton acceleration and injection into the interstellar medium on timescales of years.

\begin{figure}
\includegraphics[width=\columnwidth]{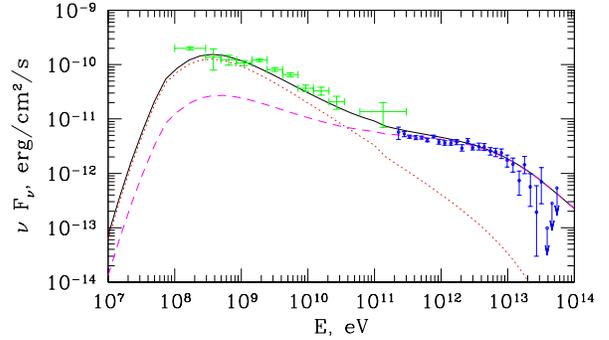}
\caption{Combined \fermi (green points) and HESS (blue points) explained by superposition (black solid line) of a proton flare of 10 years duration happened 300 years ago (magenta dashed line) and a constant source that switched on
$10^4$ years ago (red dotted line). See Section \ref{sec:discussion} for model parameters.
\label{spe} }
\end{figure}

 In Figure \ref{spe}  we compare the spectra of gamma-ray emission
resulting from realization of three different scenarios:  (i)  a proton flare of 10 years
duration that occurred 300 years ago, (ii)  a constant source that switched on
$10^4$ years ago, and (iii)  a proton flare on top of the constant source, namely the
superposition of (i) and  (ii).  To fit the data, we took the size of the gamma-ray emission region to be $R=8$~pc,
parameters of the diffusion coefficient $\beta=0.65$, $\kappa=1$, and initial proton spectrum of
$Q(E)\propto E^{-2}\exp(-E/100$~TeV$)$, with  the proton injection rate
of $1.9\times 10^{39}$erg/s for the constant source and $1.9\times 10^{42}$erg/s for the flare.

 For  this parameter set, the 300 year old flare  cannot have a strong impact on the observed TeV spectrum at this point in time, since most of the high energy protons from the flare have already escaped. On the other hand, the emission at GeV energies is produced by protons from the flare which are still trapped by diffusion in the gamma-ray production region.
To explain the TeV data we need much slower  diffusion or a fresh  injection of  protons, for example contributed
from a very recent flare, or  by the quasi-steady component of  protons. Actually the form of TeV emission doesn't depend on the age of the source if it exceeds $t_{esc}=R/c\sim 30$years.  The case of superposition (solid line)
of the flare  (dashed line) and persistent (dotted line) components of protons is shown in Figure  \ref{spe}. For the
chosen parameters, the GeV energy range of gamma-rays is dominated by the flare component of protons, while the
TeV gamma-rays  are  contributed mainly by protons from the persistent component.

Thus,  we are able to reproduce the observed  broad-band spectrum
of gamma-rays in different ways. In all cases the required injection rate is well below the Eddington limit of
the $4\times 10^6 M_\bigodot$ black hole  in GC,  $L_{Edd}=\frac{4\pi GMm_p}{c\sigma_T}\simeq 6\times 10^{44}$~erg/s.
The total energy required in  relativistic protons  currently trapped  in the gamma-ray production region  varies from $10^{49}$ to $10^{51}$~erg for different models. This energy can be injected in very different ways: in reality there has probably been a series of flares with different energetic signatures occurring throughout the life-time of the central source.


 The observed spectral and temporal properties of the GC at various wavebands are not enough to constraint all the parameters in our model. Additional information can be extracted, in principle, from the gamma-ray morphology of the inner arcminute region: above, we showed that different set of parameters describing the observed spectral properties result in very different radial profiles (e.g., Figure \ref{brightprof}).  However, with the angular resolutions of the current
 space- and ground-based detectors, we cannot distinguish between the
 different radial profiles.  Fortunately, such information can be recovered by observations of synchrotron emission
 of secondary electrons from decays of charged pions, accompanying the production of gamma-rays from
 decays of neutral muons.  Since through this channel  the electrons and gamma-rays are produced with
similar energy distributions, we can connect directly the frequency of synchrotron photons of secondary
electrons with the energy of  the ``genetically" connected gamma-rays:
\begin{equation}
\epsilon \simeq  100  \left( \frac {B} {10^{-4} \mbox{~G}}\right) \left({E_\gamma} \over {\mbox{1 GeV}} \right)^2 \mbox{MHz},
\end{equation}
where the magnetic field is  normalized to the probable minimum value  expected in the region \citep{Crocker2010}.

Thus, in the first approximation,  the morphology of the synchrotron radiation of "hadronic" origin
should be similar to the morphology of  GeV gamma-rays. While at sub-GHz frequencies, GC radio photons are attenuated by free-free absorption in dense H{\sc ii} regions between the GC and the Earth
(see \citealt{Crocker2010}),  the synchrotron emission at $\sim$ GHz frequencies arrives without
significant attenuation.  The emissivity of synchrotron radiation  of  secondary electrons in the inner few pc of the
Galaxy has been studied by  \cite{ballyntyine_synch} , based on  a model where the interactions of protons, diffusing away from  an assumed central source, supply the observed, point-like TeV signal \citep{Ballantyne07}.  They then compared the predicted synchrotron emission to the GHz radio frequency spectrum, and found that essentially all the diffuse non-thermal
GHz radio emission from the central parsecs of the Galaxy could be explained as due to emission of
secondary electrons (and positrons).  Therefore we  anticipate that the  new \fermi data
combined with  available radio measurements, should allow us
to constrain significantly the parameter space
of models positing that the GeV and TeV gamma-ray
emission of the GC is due to  hadronic interactions in the central few parsecs of GC.  Analysis of the morphology of radio emission holds out particular promise here.
 The results of such an analysis are beyond the scope of this paper and will be presented elsewhere.

It should be noted that cosmic ray electrons  produce photons not only in the radio domain;
their bremsstrahlung emission can be an important source of high energy gamma-rays. Comparing
the synchrotron and bremsstrahlung cooling times:
\begin{eqnarray}
&&t_{synch}\simeq 4\times 10^{12}\left(\frac{B}{10^{-5}\ \mbox{G}}\right)^{-2}\left(\frac{E_e}{1\mbox{~TeV}}\right)^{-1} \mbox{s}\\
&&t_{bremms}\simeq 1.2\times 10^{12}\left(\frac{n_H}{1000 \ \mbox{cm}^{-3}}\right)^{-1} \mbox{s},
\end{eqnarray}
we conclude that bremsstrahlung losses dominates at energies below:
\begin{equation}
E_e \lesssim 30\left(\frac{n_H}{10^3 \ \mbox{cm}^{-3}}\right)\left(\frac{10^{-4}\mbox{G}}{B}\right)^2\mbox{~GeV} .
\end{equation}
This suggests  that bremsstrahlung will dominate at energies less than  $\sim30$~GeV for fiducial $n_H$ and $B$ values.

 The relative importance of electron bremsstrahlung  in producing the observed gamma-ray emission, is characterized  by the ratio of cooling times of electrons and protons associated with the bremsstrahlung
and  neutral pion production, respectively: $q_{\gamma}^{br}/q_{\gamma}^{\pi^0}\simeq  (3t_{pp}/t_{br})f\simeq 4f$, where $f=n_e/n_p$ is the electron to proton energy  density ratio  \citep{aharonian_book}.
Hence, if the ratio of protons to electrons is $\gg 1$, then $\pi^0$-decay gamma-rays dominate
over bremsstrahlung.  The contribution of electron bremsstrahlung to TeV gamma-rays   is quite limited
because of  the severe energy losses of very high energy  electrons due to the synchrotron and IC losses.

Finally, we note that the \fermi data presented here  can not be explained  by  IC models proposed by \cite{atoyan_dermer04}  and \cite{Hinton07}. While these models are in a good agreement with HESS data,
they predict that the energy flux in the GeV part of the spectrum should be smaller than in TeV range,
apparently contrary to the \fermi observations.

 \section{Summary}\label{sec:summary}
We have analyzed 25 months of \fermi data on the GC region.  The \fermi LAT source 1FGL J1745.6-2900 lies within the error box of HESS source J1745-290. We found that, while below 5~GeV, the spectrum of  1FGL J1745.6-2900 has a photon spectral index similar to the HESS source, the spectrum at higher energies is better described by a steeper spectral index. We have formulated a model which produces a photon spectrum that can naturally explain the observed broad-band gamma-ray emission. This model considers the  hadronic interactions of relativistic protons which, having diffused away from a central source, presumably the central black hole, fill the inner few parsecs of our Galaxy. We have explored the parameter space of our model, in terms of the geometry, characteristics of the diffusion coefficient, and  injection rate  history.

We have shown that the available spectral information can be well described with different sets of model parameters and that additional information is required to distinguish model scenarios. Such information could be obtained from the spatial distribution of the observed gamma-ray emission; however, the required arc-second resolution cannot be reached by gamma-ray telescopes. Luckily, synchrotron emission from the secondary electrons and positrons expected in our model may be detected by radio telescopes which possess an angular resolution high enough for the purposes of distinguishing between model parameters.

\acknowledgments
The authors are grateful to Venya Berezinsky for the discussions on the transition from diffusion to rectlinear regime. The authors wish to acknowledge
the SFI/HEA Irish Centre for High-End Computing
(ICHEC) for the provision of computational facilities and support.
The work of D.M. was supported by
grant 07/RFP/PHYF761 from Science Foundation Ireland (SFI)
under its Research Frontiers Programme.

\bibliography{GCem}
\end{document}